\documentclass[12pt,a4paper]{iopart}
\usepackage[pdfencoding=auto, psdextra]{hyperref}
\usepackage{hyperref}
\usepackage{harvard}
\usepackage{graphicx}
\usepackage{subfigure}
\usepackage{bbm}
\usepackage{mathrsfs}

\newcommand\de{\delta}
\newcommand\ep{\epsilon}
\newcommand\ze{\zeta}

\renewcommand\th{\theta}

\newcommand\rh{\rho}

\newcommand\Ga{\Gamma}

\newcommand\Ph{\Phi}

\newcommand\<{\langle}
\renewcommand\>{\rangle}

\newcommand\ie{\emph{i.e.}}

\newcommand\beq{\begin{equation}}
\newcommand\eeq{\end{equation}}
\newcommand\bea{\begin{eqnarray}}
\newcommand\eea{\end{eqnarray}}
\newcommand\bal{\begin{align}}
\newcommand\eal{\end{align}}

\newcommand\fr{\frac}

\renewcommand\bal{\mbox{\boldmath$\alpha$}}

\begin{document}

\title{Scale invariant curvature perturbations from a spontaneously decaying scalar field}
\author{R. Lieu$^1$, C.-H. Shi$^2$}
\address{$^1$ Department of Physics and Astronomy, University of Alabama, Huntsville, AL 35899.}
\address{$^2$ Chinese Academy of Sciences South America Center for Astronomy,\\
\quad China-Chile Joint Center for Astronomy (CASSACA), Camino El Observatorio  1515, Las Condes, Santiago, Chile}
\ead{E-mail:lieur@uah.edu}

\vspace{10pt}
\begin{indented}
\item[]\today
\end{indented}

\begin{abstract}
The evolution of superhorizon curvature perturbations in a two-component interacting universe is considered.  
It is found that adiabatic modes conserve the total curvature perturbation $\ze$, {\it unless} there are stages in which the rate of dissipation of one component into another is not constant. 
Moreover, our result shows that when the rate is varying it {\it is} possible for 'isocurvature' perturbations generated during reheating to alter the amplitude of an adiabatic curvature mode even when the mode  is outside the horizon. Specifically, if an indefinitely large rate $\Gamma$ for massive particles decaying into photons develops rapidly amid vanishingly small initial values (before decay) of the total curvature $\zeta_i$ and Newtonian potential $\Phi_i$, such that the product $\Gamma\zeta_i$ and $\Gamma\Phi_i$ become a pair of finite and universal constants for all superhorizon scales afterwards, Harrison-Zel'dovich scale-invariant power spectrum could be synthesized from a homogeneous state without inflation at all.
\end{abstract}

\vspace{2pc}

\submitto{\CQG}

\maketitle

\section{Introduction}

In cosmological inflation, \cite{Guth1981,Linde1982,Albrecht1982,Starobinsky1987,Allahverdi2010,Lalak2007,Lozanov2020},
it is widely believed 
a single scalar field drives the inflation, which is compatible with the observational data \cite[for example]{Chowdhury2019}.
Although models with multi-field drew some attention \cite[for instance]{Abedi2019,Martin2021} and are not ruled out by any observational data, our discussion here only limits to the single-field mode, which states
that there exist many e-folds of expansion of space during the slow-roll phase of the process, by the end of which the inflaton scalar field $\phi$ rolls down a steep slope of the $V(\phi)$ versus $\phi$ curve towards the bottom of the potential well $V(\phi$), where it is trapped and the ratio $w_v$ of its pressure $p_v$ to energy density $\rh_v$ undergoes rapid coherent oscillations at the frequency $\sim V''(\phi_{\rm min})$.   Specifically, the two quantities are given to the lowest order by 
\beq 
p_v = \dot\phi^2/2 - V(\phi),~\rh_v = \dot\phi^2/2 + V(\phi); \label{energy} 
\eeq
so that $w_v=1$ at the potential minimum where $V (\phi_{\rm min})=0$ while $w_v=-1$ at the two topmost points of each cycle where $\dot\phi =0$.  In the standard treatment of the problem, this oscillation is regarded as a `massive condensate' of average $\< w_v \> = 0$, \ie~cold non-relativistic particles capable of decaying into radiation at some constant rate $\Ga$ that becomes finite only when the inflaton is trapped at the potential minimum, a phase representing the final leg of inflation known as reheating, \cite{Bardeen1983}.   The resulting outcome is the conservation of the total superhorizon curvature perturbation in the longitudinal gauge, a quantity often symbolized by $\ze$
($\zeta$ is in fact {\it gauge invariant}).

There are two fundamental limitations to the above approach to the evolution of $\ze$.  First is the procedure of averaging $w_v$, which is valid only if the oscillation of $w_v$ does not cause $\ze$ to become large at certain times.  Second is the neglect of the role of $\dot\Ga$.  If the decay rate $\Ga$ is vanishingly small during slow-roll, but becomes significant in reheating, there must have been an intervening stage of finite $\dot\Ga$, and the question is whether this too would exert any appreciable effect upon $\ze$.

Let us remind the reader of the relevant background material, \cite{LiddleLyth2000}.  The total curvature perturbation, defined as 
\cite{MWU2003}
\beq 
\ze= -\Ph - H\fr{\de\rh}{\dot\rh}, \label{total} 
\eeq 
may also be expressed in terms of the contribution from the multiple fluids that constitute the universe as 
\beq \ze = \sum_{\nu} \fr{\dot\rh_\nu}{\dot\rh}~\ze_\nu, \label{ze} \eeq
where 
\beq 
\ze_\nu = -\Ph - H\fr{\de\rh_\nu}{\dot\rh_\nu}. \label{zenu} 
\eeq   
In both equ.(\ref{total}) and (\ref{zenu}) the $-$ sign in front of $\Ph$ is the convention of Malik et al.(2003).

If each fluid has an `adiabatic' sound speed {\it and} constant equation of state, or more precisely if the derivative $dp_\mu/d\rh_\mu$ of fluid $\mu$ happens to equal the following quantity  
\beq 
\fr{\dot p_\mu}{\dot\rh_\mu} = \fr{w_\mu \dot\rh_\mu}{\dot\rh_\mu} = w_\mu, \label{ssp} 
\eeq 
as shown in equ.(\ref{DotZeta1}) in \ref{appendix:MathSupplement},
the rate of change of $\ze$ will be
\beq 
\dot\ze = \fr{-1}{6\dot\rh(\rh+p)} \sum_{\mu,\nu} \dot\rh_\mu \dot\rh_\nu (w_\mu - w_\nu) S_{\mu \nu}, \label{dotze1} 
\eeq 
where 
\beq 
S_{\mu \nu} = 3(\ze_\mu - \ze_\nu) \label{ent} 
\eeq 
is the relative entropy (isocurvature) perturbation between a pair of fluids.
In the present context of the {\it two} interacting fluids of vacuum (the inflaton) and radiation, {\it viz.} $v$ and $r$, equ.(\ref{dotze1}) reduces to 
\beq 
\dot\ze = -\fr{H}{\dot\rh^2} \dot\rh_v \dot\rh_r (w_v - w_r) S_{vr}, \label{dotze2} 
\eeq  
where  
\beq 
\dot\rh = -3H[(1+w_v)\rh_v + (1+w_r)\rh_r]= -3H(\rh+p) \label{trates} 
\eeq 
is assumed.

Two points about equ.(\ref{dotze1}) are noteworthy.  (a) The growth equation for $\ze$ is {\it only} a function of the {\it difference} between $\ze_\mu$ and $\ze_\nu$.  (b) There is no evolution in this difference unless an energy-exchange interaction takes place between the two fluids.  
Thus, if the inflaton decays to radiation at some constant rate $\Ga$, one may describe the change in the mean density of the two fluids by the equations 
\numparts
\beq 
\dot\rh_v = -3(1+w_v)H\rh_v- \Ga\rh_v,\label{rates_v}
\eeq
\beq
\dot\rh_r = -3(1+w_r)H\rh_r + \Ga\rh_v, \label{rates_r} 
\eeq 
\endnumparts
in which case the relative entropy $S_{vr}$ of the two-component fluid evolves, as derived in \ref{appendix:MathSupplement}, as  
\beq 
\dot S_{vr} = \Ga\left[\fr{\rh_v}{2\rh} \fr{\dot\rh_v}{\dot\rh_r} \left(1-\fr{\dot\rh_r^2}{\dot\rh_v^2}\right) - \fr{\dot\rh_v}{\dot\rh_r}\right]S_{vr}. \label{Seq} 
\eeq  
Now neither (a) nor (b) is valid in general.  If the equation of state of either fluid, characterized by $w_v$ and $w_r$, is time-varying, it will be shown that (a) no longer holds.  Similarly, if the $v \to r$ decay rate is not constant, (b) will need modification.

\section{Generalized Growth Equations}
\label{sec::growthequation}

Let us first extend equ.(\ref{dotze1}) to include the possibility of finite $\dot{w_v}$.  
We start by taking the time derivative of equ.(\ref{total}) and using (\ref{trates}), to get 
\beq 
\fl
\dot\ze = -\dot\Ph +
\fr{\de\dot\rh}{3[(1+w_v)\rh_v + (1+w_r)\rh_r]} 
-\fr{(\de\rh_v + \de\rh_r)[(1+w_v)\dot\rh_v + (1+w_r)\dot\rh_r + \rh_v\dot w_v]}{3[(1+w_v)\rh_v + (1+w_r)\rh_r]^2}.
\label{dotze3} 
\eeq 
Next, we need the following equation 
for the evolution of $\rh$,
which is the perturbed equation of energy conservation,
\begin{equation}
\fl
\eqalign{
\de\dot\rh & = 3(\rh + p)\dot\Ph -3H(\de\rh+\de p) \cr
& = 3[(1+w_v)\rh_v + (1+w_r)\rh_r]\dot\Ph -3H\left[\left(1+w_v+\fr{\dot w_v\rh_v}{\dot\rh_v}\right)\de\rh_v + (1+w_r)\de\rh_r\right].
}\label{dotdrh} 
\end{equation}  
Note that in establishing equ.(\ref{dotdrh}) we invoked an `adiabatic' sound speed amidst a time-varying equation of state for the inflaton, {\it viz.}~by `adiabatic' one means that \beq \fr{dp_v}{d\rh_v} = \fr{\dot p_v}{\dot\rh_v}, \label{adiabatic} \eeq with the latter of the form  $\dot p_v/\dot\rh_v = (w_v \dot\rh_v + \dot w_v \rh_v)/\dot\rh_v$ in the case of a finite $\dot w_v$.  The only difference from the subtext of equ.(\ref{dotze2}) as clarified by the words before that equation is the absence of time dependence in the $w_v$ there.  
It has to be emphasized that 
the quantity $c_\mu^2$ is {\it defined} as $\dot p_\mu/\dot\rh_\mu$ in many literatures \cite[for instance]{MWU2003,Riotto2003}.
Thus, in the present paper here, the $c_\nu^2$ above defined 
is {\it not} automatically the square of the actual sound speed of fluid $\mu$ (\ie~it is not $dp_\mu/d\rh_\mu$); only when the sound speed is `adiabatic' will the two be equal.  
Substituting equ.(\ref{dotdrh}) into (\ref{dotze3}), one arrives at 
\beq 
\dot\ze = \fr{3H}{\dot\rh} [(w_v-w_r) \dot\rh_v + \rh_v\dot w_v] (\ze-\ze_v), 
\label{dotze4} 
\eeq 
after laborious algebra.

Next, we turn to $\dot\Ga$.   Here we revisit the first principles, by returning to equ.(\ref{DotZetaMu})
and invoking equ.(\ref{adiabatic}) for {\it each} fluid;  for the inflaton this is the so-called `canonical inflation' scenario (as opposed to the brane inflation scenario of subluminal sound speed advocated \cite{Peiris2007}).  
This enables the $\de p_{\rm intr}$ term of (2.33) to vanish.  For the two interacting fluids $v$ and $r$ as in equ.(\ref{trates}), (\ref{rates_v}) and (\ref{rates_r}), the remaining terms of (2.33) lead, after some algebra and with the aid of (2.37), to the following equation,
$$ \dot\zeta_v = -\fr{H}{\dot\rh_v} \left(\de Q_v - \fr{\dot Q_v}{\dot\rh_v} \de\rh_v\right) - \left(\fr{HQ_v\dot\rh}{2\rh\dot\rh_v}\right) \left(\fr{\dot\rh_r}{\dot\rh}\right) \left(\fr{\de\rh_v}{\dot\rh_v} - \fr{\de\rh_r}{\dot\rh_r}\right);  $$ or, in alignment with the (2.33) and (2.36) combination, \beq \dot\zeta_v = -\fr{H}{\dot\rh_v} \left(\de Q_v - \fr{\dot Q_v}{\dot\rh_v} \de\rh_v\right) + \fr{Q_v}{6\rh\dot\rh_v} \dot\rh_r S_{vr}. \label{dotzeka} \eeq  A similar equation for $\dot\ze_r$ exists, with the $v$ and $r$ subscripts interchanged.

Since the inflaton and radiation interact with each other in the manner of equ.(\ref{rates_v}) and (\ref{rates_r}), we have $Q_v = -\Ga\rh_v$, $Q_r=\Ga\rh_v$.  
Equ.(\ref{dotzeka}) and its counterpart may now be written as 
\numparts
\begin{eqnarray}
\dot\ze_v & = & -\fr{\Ga\rh_v\dot\rh_r}{2\rh\dot\rh_v} (\ze_v - \ze_r) + \fr{\dot\Ga\rh_v}{\dot\rh_v} (\ze_v + \Ph); \label{pairA}\\
\dot\ze_r & = & \left(-\fr{\Ga\rh_v\dot\rh_v}{2\rh\dot\rh_r} + \fr{\Ga\dot\rh_v}{\dot\rh_r}\right) (\ze_v - \ze_r) -\fr{\dot\Ga\rh_v}{\dot\rh_r} (\ze_r + \Ph),  \label{pairB} 
\end{eqnarray}
\endnumparts
where use was made of equ.(\ref{zenu}) and (\ref{ent}) to arrive at the $\dot\Ga$ terms.  There is moreover an accompanying equation, the superhorizon version of the gravitational Poisson equation, or equ.(\ref{dotPhi}), which yields (namely. equ.(\ref{dotPhi2}))
\beq 
\dot\Ph + H\Ph = \fr{\dot\rh_v (\ze_v + \Ph) + \dot\rh_r (\ze_r + \Ph)}{2\rh}. \label{Poisson} 
\eeq
In the $\dot\Ga = 0$ limit, one can use only equ.(\ref{pairA}) and  (\ref{pairB}) in conjunction with equ.(\ref{ent}) to obtain (\ref{Seq}).

But as far as the rapid variations in 
$\ep=1+w_v
=\dot{\phi}^2/[\dot{\phi}^2/2 + V(\phi)]$, 
($\dot w_v = \dot\ep$) and $\Ga$ during reheating are concerned, we prefer the perturbation variables 
$(\ze, \ze_v, \Ph)$ to $(\ze_v, \ze_r, \Ph)$.   
This new set will prove to be a better choice because $\dot\rh_r$ vanishes at the end of reheating (when it changes sign from positive to negative as the inflaton decay wanes and the radiation is redshifted by the Hubble expansion) and causes the coefficients of the growth equations for $\ze_r$, equ.(\ref{pairB}), to diverge.  In terms of the new variables, equ.(\ref{pairA}) becomes, with the help of equ.(\ref{ze}), 
\beq 
\dot\ze_v = \fr{\Ga\rh_v\dot\rh}{2\dot\rh_v\rh} (\ze - \ze_v) + \fr{\dot\Ga\rh_v}{\dot\rh_v} (\ze_v + \Ph).  
\label{npair}  
\eeq
Together with equ.(\ref{dotze4}) and (\ref{Poisson}), with the latter re-expressed in terms of the new variables as 
\beq 
\dot\Ph + H\Ph = \fr{\dot\rh}{2\rh} (\ze + \Ph),  \label{nPoisson} 
\eeq 
one has a complete set of three equations and three unknowns, as before.
One has a complete set of three equations (\ref{dotze4}), (\ref{npair}), and (\ref{nPoisson}) and three unknown variables of perturbation $\zeta$, $\zeta_v$, and $\Phi$ as before.

\section{Reheating at the end of inflation}
\label{sec::reheating}
It is now necessary to be very specific about the model.  We assume that the time $t_h$ at which the inflaton scalar field departs from slow-roll and enters the coherent oscillation phase is sharp and well-defined, and that $\Ga$ becomes finite, \ie~$\Ga > 0$ at this time  (the physical statement of the `sudden turn on' approximation for $\Ga (t)$ at $t=t_h$ is actually $\dot\Ga \gg \omega_h \sim V''(\phi_{\rm min})$, the frequency of the oscillation).  Mathematically we write 
\numparts
\begin{eqnarray}
\Ga (t) & = & \Ga \th (t-t_h); \label{Ga}\\ 
\ep (t) & = & 
1 - \cos [\omega_h (t-t_h) + \sqrt{2\ep_h}\,], 
\ \omega_h \gg \Ga
\gg H,
\label{eposc} 
\end{eqnarray}
\endnumparts
where $\th (t)$ is the Heaviside step function, and $\ep_h = \ep(t_h)$ is a positive constant $\ll 1$.  
In the construction of equ.(\ref{eposc}), $\ep (t)$ is small and oscillates between $0$ and $2$ at $t > t_h$
(because $\epsilon=1+w_v$, with $-1\leq w_v\leq 1$ for all reasonable equation of state).
$\Ga(t)$ is lately discussed in a form of a continuous function, such as the sigmoid function \cite{Diaz2020}, which described a high temperature, within the range of Plank's bounds, during the reheating.

Our period of interest is only restricted to $t\geq t_h$, however, since $\Ga =0$ and hence $\rh_r = \dot\rh_r =0$ for $t<t_h$ (there is no radiation in the absence of dissipation), \ie~we have $\ze = \ze_v$ by equ.(\ref{ze}) and $\ze =$~constant by equ.(\ref{dotze4}).  
Moreover, even when $t> t_h$ one could 
(but erroneously)
argue by appealing to equ.(\ref{Ga}) that the situation will not change, because equ.(\ref{dotze4}) and (\ref{npair}) can be combined to yield 
\beq
\fl
\dot\ze - \dot\ze_v = -\left[(\frac{4}{3}-\ep)\fr{(3\epsilon H+\Ga)\rh_v}{\ep\rh_v + \frac{4}{3}\rh_r} +\fr{\rh_v}{\ep \rh_v + \frac{4}{3} \rh_r} \dot\ep  + \fr{\Gamma(3\ep\rh_v + 4\rh_r)}{2\rh(3\epsilon + \Gamma/H)}\right](\ze - \ze_v);~{\rm for}~t>t_h, 
\label{coherent} 
\eeq 
and the absence of any $\dot\Ga$ term in equ.(\ref{coherent}) allows one to deduce that $\ze = \ze_v$ is maintained at $t>t_h$ as in the case of $t< t_h$.  
The flaw in using this line of reasoning as the basis for the conservation of $\ze$ is that it neglects what happens at $t=t_h$.  Here, by equ.(\ref{Ga}) $\dot\Ga (t) = \Ga \de (t-t_h)$ becomes large, and so the $\dot\Ga$ term of equ.(\ref{npair}) dominates, leading to a finite jump in $\ze_v$ {\it without} an accompanying jump in $\ze$ (the latter is because in equ.(\ref{dotze4}) $\dot\ze$ does not contain any $\dot\Ga$ term).  
Thus, the correct conclusion is that as a result of $\ze = \ze_v$ before $t=t_h$, one {\it must} have $\ze \neq \ze_v$ afterwards.  By equ.(\ref{dotze4}), this phenomenon causes $\ze$ to evolve during times $t>t_h$.

It is therefore important to calculate the change in $\ze_v$ across $t=t_h$. 
Here the reader is referred to \ref{appendix:SolutionReheating} for details on the exact jump condition.
Moreover, \ref{appendix:PostReheating} provided an analytic solution for $\zeta(t)$ at early and late times in the post-reheating era.

\section{Numerical Solution}
\label{sec::numerical}
While the analytical treatment of \ref{appendix:SolutionReheating} and C can shed light on the behaviour of the perturbation variables $\zeta$, $\zeta_v$, and $\Phi$ during specific (and critical) moments of time, one must resort to numerical methods in order to evolve these variables for $t=t_h^-$ to late time $t\gg1/H$.

\begin{figure}
    \centering
    \subfigure[]{\includegraphics[width=0.43\textwidth]{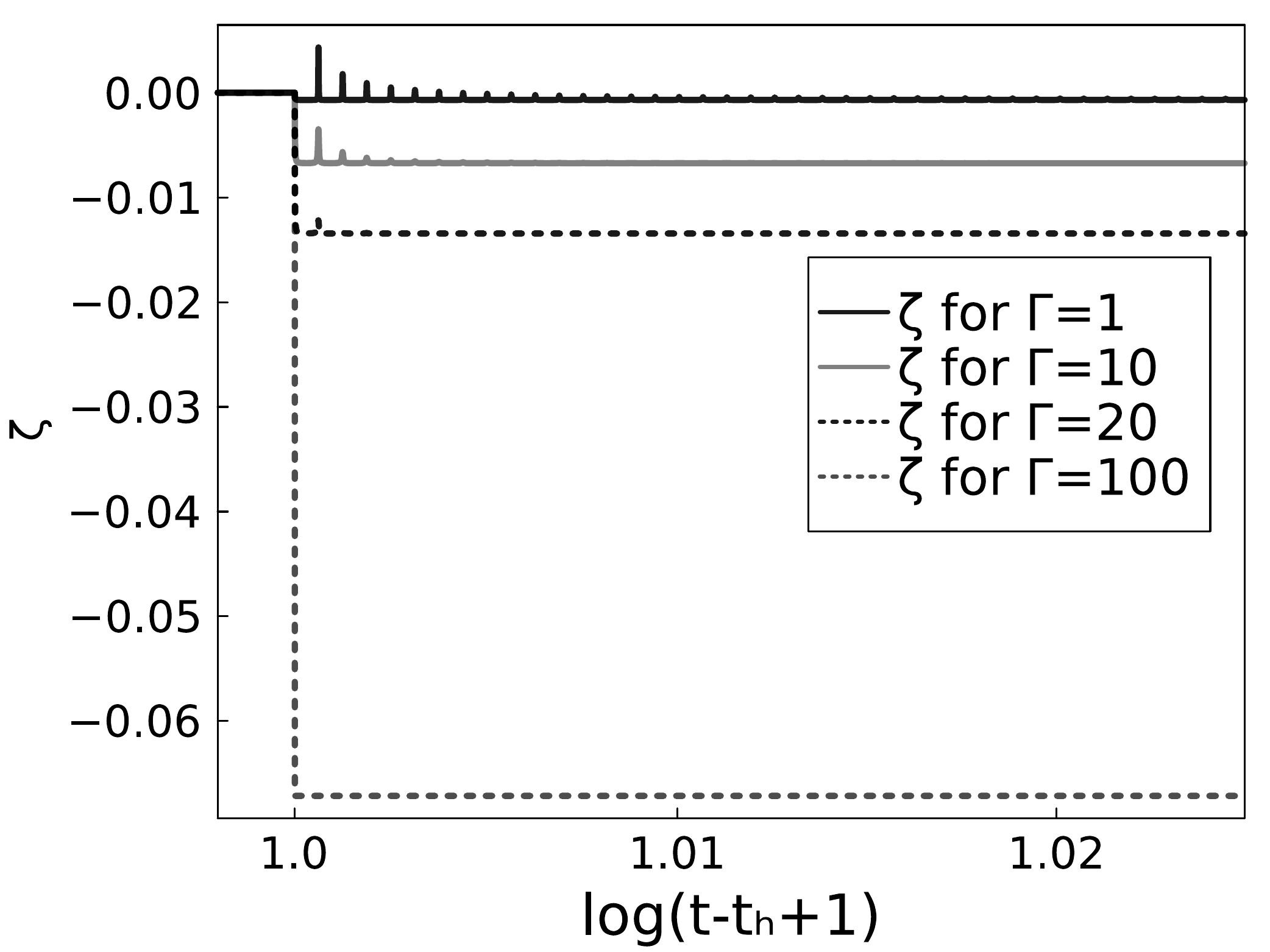}\label{fig:a}}
    \subfigure[]{\includegraphics[width=0.43\textwidth]{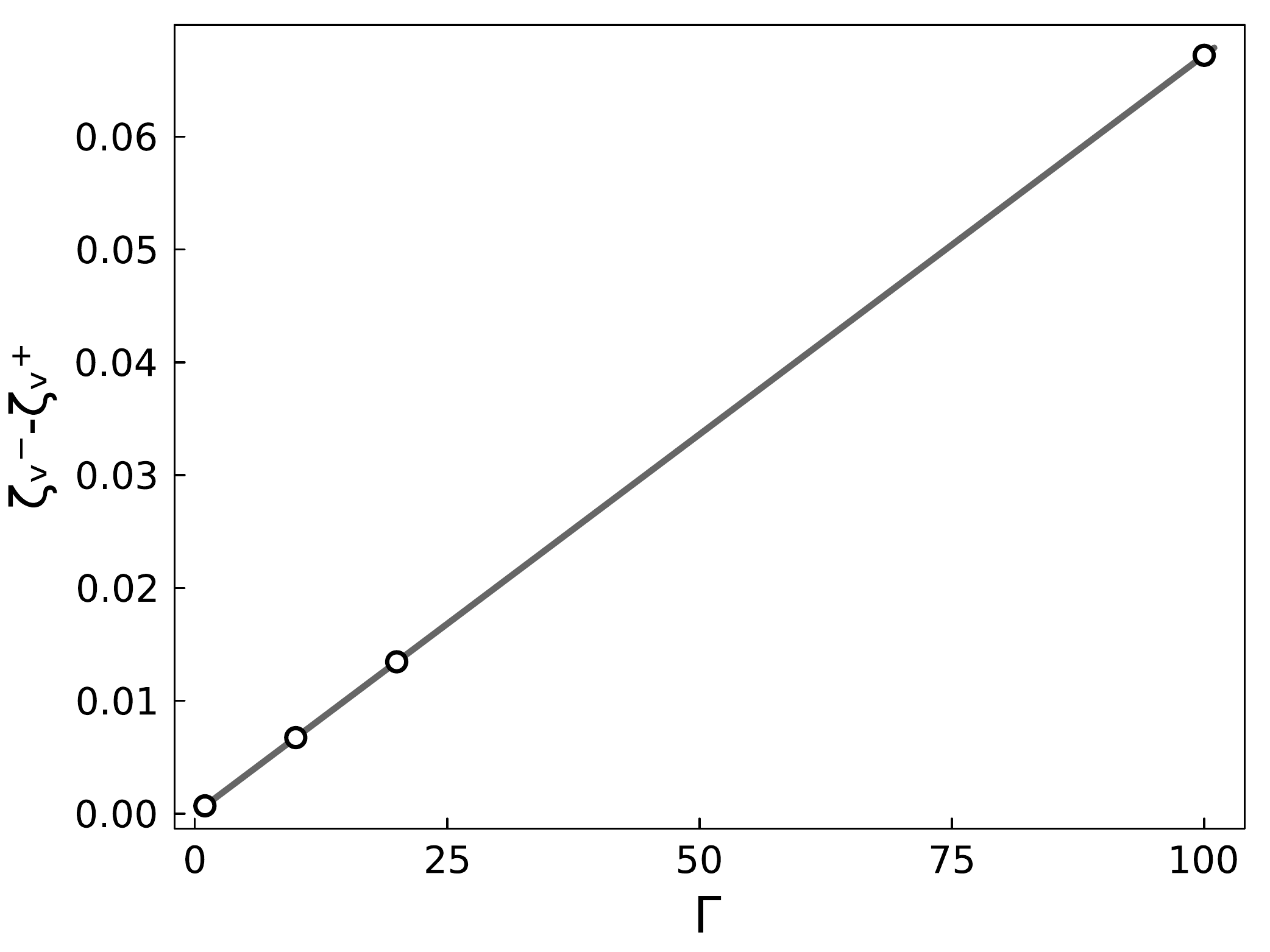}\label{fig:c}}
    \subfigure[]{\includegraphics[width=0.9\textwidth]{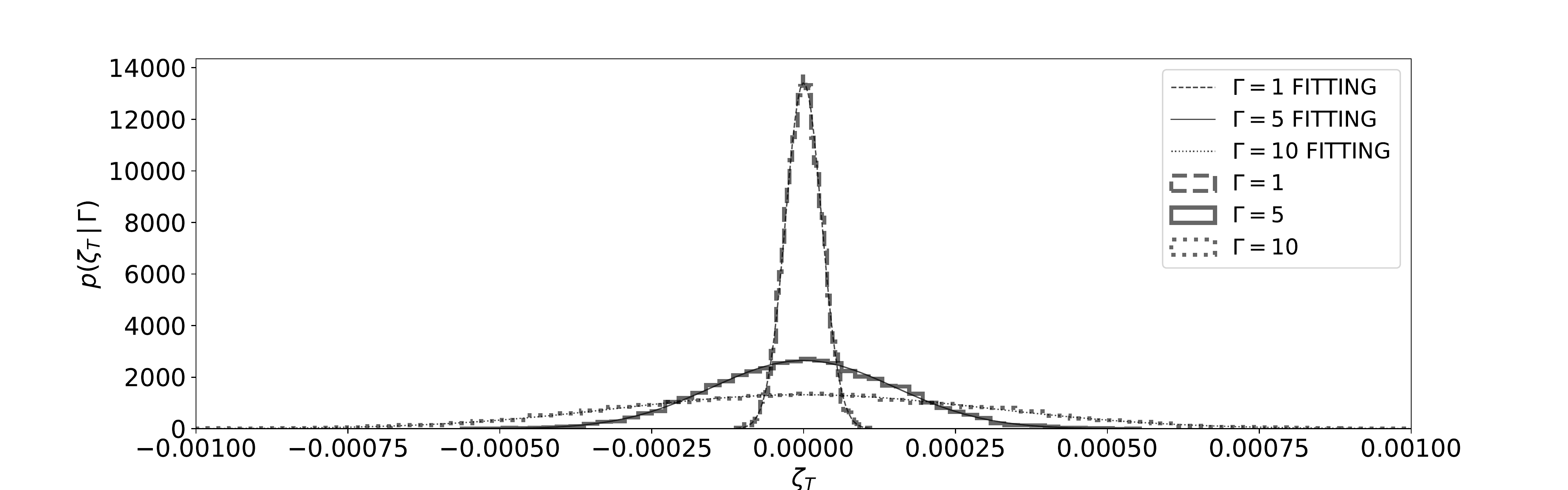} \label{fig:d}}
    \caption{
    {\bf(a)} Reheating (large decay rate) commences as a step function $\Gamma\theta(t-t_h)$ and triggers a jump in $\zeta_v$. The ensuing jump in $\zeta$ is actually gradual. 
    The subsequent spikes are due to coherent oscillations in $\epsilon(t)=1+w_v(t)$. 
    This subfigure clearly shows how adiabatic curvature perturbations are not conserved during reheating because of $\dot\Gamma$. It means the superhorizon isocurvature modes, sourced by reheating, can affect the adiabatic modes. 
    The $\zeta$-axis is plotted in linear scale. The horizontal variable is $t-t_h+1$ in order to plot this axis in log-scale and to show spikes at $t$ close to $t_h$.
    After $t=t_h$, $\zeta$, instead of being discontinuous, drops steeply and approaches $\zeta_v$ rapidly. The greater $\Gamma$ is, the deeper $\zeta$ drops.
    Those peaks, standing on solutions of $\Gamma=1$, $10$, and $20$, are spikes described in equ. (\ref{coherent3}). When $\Gamma=100$, those spikes are too small to be seen in the figure.   
    {\bf (b)} 
    Show the approximate proportional relation between ($\zeta_v^--\zeta_v^+$) and $\Gamma$. 
    The heights of $\zeta_v^--\zeta_v^+$, shown as open circles, versus values of $\Gamma$ corresponding to (a). $\zeta_v^--\zeta_v^+$ is proportional to $\Gamma$ approximately. The slope of the black straight line is $[\zeta_v^-+\Phi(t_h)]\eta/2$ (see \ref{appendix:SolutionReheating}).
    {\bf (c)} To test whether the Gaussianity is distorted by reheating,
    $10^4$ samples of $\zeta_0$ are drawn from a Gaussian distribution of $\mathcal{N}(0.0, \sigma_{\zeta0})$. And $10^4$ of $\Phi_0\sim\mathcal{N}(0,\sigma_{\Phi0})$ are drawn independently. With a given $\Gamma$, one can calculate numerically the values of $\zeta_T=\zeta(t\gg1)$. These $\zeta_T$ form a histogram. 
    The histogram is drawn in the probability density, in which each bin displays its density $=($raw count$)/[($total count$)*($bin width$)]$.
    The histogram fits a Gaussian model $p(\zeta_T|\Gamma)=$
    $\mathcal{N}(0,\sigma_T)$ very well. When varying the value of $\Gamma$, one observes that the Gaussian profile of $\zeta_T$ changes accordingly with $\sigma_T$ approximately proportional to $\Gamma$.}
    \label{fig}
\end{figure}

We combine equ.(\ref{dotze4}), (\ref{npair}), and (\ref{nPoisson}) to explore the evolutions of variables $\zeta$, $\zeta_v$ and $\Phi$ as functions of time $t$. Because of the presence of $\rho_v$ and $\rho_r$ in these equations, equ.(\ref{rates_v}) and (\ref{rates_r}) have to be included in the above combination. 
To be convenient, 
time is in the unit of Hubble time as setting $H$ to be $1$, and the time variable $t$ is replaced with $t-t_h$ in the numerical computation. (To avoid confusion, this replacement is constrained only to the numerical code. In the context of the paper, including the figures, $t-t_h$ remains.)
Then we have a combination of five {\it ODEs} (ordinary differential equation), which is consisted of (\ref{rates_v}), (\ref{rates_r}), (\ref{dotze4}), (\ref{npair}), and (\ref{nPoisson}), 
Runge-Kutta\footnote{To be specific, we use algorithm Verner's $7/6$ Runge-Kutta method here.} 
method is employed to solve our ODEs.  
In the stage of 
$t<t_h$,
the radiation field is absent before reheating happens. There is $\rho_{v0} =1$ and $\rho_{r0}=0$. Perturbation only exists in the curvature, which gives us $\zeta_0=\zeta_{v0}$. We choose the initial value of the perturbation and potential are $\zeta_0=\zeta_{v0}=3.0\times10^{-5}$ and $\Phi_0=1.0\times10^{-5}$, respectively.  Integrating the combination of ODE from 
$t-t_h=-0.5$ to $0$, we obtain the evolution of the perturbation before reheating. 
With the aid of the jump condition, derived in \ref{appendix:SolutionReheating}, 
we implemented the analytical solution of \ref{appendix:SolutionReheating} 
across the instance of $t=t_h$ and continue integrating ODEs forwards until reaching $t-t_h\gg1$, which is $t-t_h=50$ 
(or $500$ initial Hubble times) 
in the example of our numerical calculations here. 

Before reheating where $t<t_h$, $\zeta$ and $\zeta_v$ 
remain equal to each other, both are at the initial value of $\zeta_0$.
Reheating (large decay rate) commences as a step function $\Gamma\theta(t-t_h)$ and triggers a jump in $\zeta_v$ to another value $\zeta_T=\zeta_v(t\gg1)$ around $t=t_h$ and maintains itself at this value $\zeta_T$. 
The ensuing jump in $\zeta$ is actually gradual. 
The subsequent spikes are due to coherent oscillations in $\epsilon(t)=1+w_v(t)$. This subfigure clearly shows how adiabatic curvature perturbations are not conserved during reheating because of $\dot\Gamma$. It means the superhorizon isocurvature modes, sourced by reheating, can affect the adiabatic modes. 

One must now account for the salient features of fig. \ref{fig:a} in analytical terms.
In equ.(\ref{zthplus}), we derived the jump condition of $\zeta_v$ across $t=t_h$, due to the $\dot\Gamma$ term of equ.(\ref{npair}) under the scenario of $\dot\Gamma$ being given by equ.(\ref{Ga}). 
Since $\alpha$ changes slowly when $\Gamma$ increases, as discussed in \ref{appendix:SolutionReheating},
the height of the jump $\zeta_v^--\zeta_v^+$ is approximately proportional to $\Gamma$, shown as open circles in fig. \ref{fig:c}. The proportional factor approximates to $[\zeta_v^-+\Phi(t_h)]\eta/2$, which is the slope of the black straight line in fig. \ref{fig:c}.  
One also finds from equ.(\ref{dotze4}) and (\ref{npair}) that neither $\zeta$ nor $\Phi$ jumps across $t=t_h$. In the immediate aftermath of $t=t_h$, when $\rho=\rho_v$, (\ref{trates}), (\ref{rates_v}) and (\ref{npair}) indicate that $\dot\zeta_v\approx3H(1+w_v)(\zeta-\zeta_v)/2$. Thus $\zeta_v$ changes on timescale $\sim1/H$, which is very slow (in our numerical solution $H=1$ at $t=t_h-0.5$). 
To find out how $\zeta$ evolves during this time, return to equ.(\ref{rates_v}) and (\ref{dotze4}) to observe that the dominant contribution is from the term of $\dot\epsilon$ (since $\omega_h\gg\Gamma\gg H$), namely
\begin{equation}
\dot\zeta-\dot\zeta_v  = -\frac{\omega_h\sin[\omega_h(t-t_h)+\sqrt{2\epsilon_h}]}{1-\cos[\omega_h(t-t_h)+\sqrt{2\epsilon_h}]}(\zeta-\zeta_v).
\label{coherent2}
\end{equation}
At early time, 
$\dot\zeta-\dot\zeta_v\approx-\omega_h\sqrt{2/\epsilon_h}(\zeta-\zeta_h)$ 
is negative when $\zeta>\zeta_v$ and positive when $\zeta<\zeta_v$. Thus the sign of 
$\dot\zeta-\dot\zeta_v$ is always to reduce the difference $|\zeta-\zeta_v|$.  And because $\omega_h$ is the largest of all the frequencies, the drop of $|\zeta-\zeta_v|$ to negligible values occurs on the timescale 
$\sim\sqrt{\epsilon_h}/\omega_h\ll1/\omega_h$. 
Towards larger times $t\sim1/\omega_h$, equ.(\ref{coherent2}) may be integrated to yield 
\begin{equation}
\zeta-\zeta_v  = \frac{1-\cos(\sqrt{2\epsilon_h})}{1-\cos[\omega_h(t-t_h)+\sqrt{2\epsilon_h}]}(\zeta^+-\zeta_v^+).
\label{coherent3}
\end{equation}
Here we see there are resonances in $\zeta-\zeta_v$ when $\omega_h(t-t_h)+\sqrt{2\epsilon_h}=2n\pi$, which explains the spikes of fig. \ref{fig:a}. 
And the reason why the spikes gradually become shorter is that $\rho_r$ develops as $t$ increases beyond $t_h$. Specifically, the denominator on the right side of equ.(\ref{coherent3}) reads 
$1-\cos[\omega_h(t-t_h)+\sqrt{2\epsilon_h}] + 4\rho_r/(3\rho_v)$, as can be seen in equ.({\ref{coherent}). 
This means that the denominator becomes further and further above zero every cycle around when it is minimized. The problem is particularly acute when $\Gamma$ is larger, as more radiation $\rho_r$ is produced in the same amount of elapsed time. In fact, when $\Gamma=100$ (i.e. $\Gamma=0.01\omega_h$) the spikes are hardly visible.

The overall conclusion is, in agreement with fig \ref{fig:a}, that after $\zeta_v$ jumps, it stays constant as $\zeta$ rapidly evolves to become equal to $\zeta_v$.  Thus the excess curvature perturbation created by the $\dot\Gamma$ term at $t=t_h$ is to cause $\zeta$ to change by an amount proportional to the product of $\Gamma$ and the value of $\zeta_v + \Phi$ just before $t=t_h$. This is borne out by the fig. \ref{fig:a}.  

Reheating may affect adiabatic mode amplitudes but it does not distort the Gaussianity in $\zeta$. This means if an initial $\zeta_0$ variable is normally distributed so will the final. 
As discussed in \ref{appendix:SolutionReheating}, the standard deviation $\sigma_T$ of the Gauss distribution of $\zeta_T$ is approximately proportional to the value of $\Gamma$.
To test whether Gaussianity nature of $\zeta$ is distorted by the reheating, we drew $10^4$ samples of $\zeta_0\sim\mathcal{N}(0, ~\sigma_{\zeta0})$ and $\Phi_0\sim\mathcal{N}(0,\sigma_{\Phi0})$. Integrating them numerically to $t-t_h=50$ with a given $\Gamma$, we obtain $10^4$ corresponding results of $\zeta_T$. When setting $\sigma_{\zeta0}=1.48\times10^{-6}$ and $\sigma_{\Phi0}=1.0\times10^{-6}$, it yields that $\sigma_T=2.97\times10^{-5}$. 
Varying the value of $\Gamma$, the histogram of $\zeta_T$ changes accordingly, shown fig. \ref{fig:d}. 
It is emphasized that the histogram here is drawn in its probability density. The density on each bin is $($its raw count$)/[($total conut$)*($bin width$)]$.
$\sigma_T$ is found to assume values, with the same $\zeta_0$ and $\Phi_0$ as well as other parameters, of $1.51\times10^{-4}$ for $\Gamma=5$, $3.02\times10^{-4}$ for $\Gamma=10$, and $3.02\times10^{-3}$ for $\Gamma=10^2$ (this case is not plotted in fig.\ref{fig:d} because its Gauss profile becomes too flat to be seen).

\section{Scale Invariant Adiabatic Curvature Modes Without Inflation}
\label{sec:withoutInflation}

Guided by the results of the previous section, one could envisage a viable alternative approach to the origin of structures which bypasses inflation altogether. To elaborate, if reheating could seed adiabatic perturbations via particles decaying into radiation, it would seem more natural to envisage an initial condensate of massive particles instead of the inflaton.  This may leave the remaining problems of flatness, horizon, and absence of monopoles unsolved, but as pointed out by \cite{Lieu2013,Kibble2013}, to successfully trigger 50-60 e-folds of inflation one must ignore many even more complicating  features of the pre-inflationary state than the three problems listed above.  Thus inflation opens more difficult questions than it answers.

Let us therefore proceed to postulate the early Universe as a homogeneous distribution of massive particles, i.e. $\zeta_m\Phi\rightarrow0$ for $t<t_h$ ($\zeta_m$ now replaces $\zeta_v$; also $\dot\rho_v\rightarrow\rho_m$ and $\epsilon=1$), which at some point decays into radiation. Let the commencement of the decay be at $t=t_h$ and consists of an episode of sudden and spontaneous decay, by writing.

\begin{equation}
\dot\Gamma=\gamma_0/\tau^2,\ {\rm for}\ t_h-\tau/2\leq t\leq t_h+\tau/2;\ {\rm and}\ \dot\Gamma(t)=0\ {\rm otherwise},
\end{equation}
where $\gamma_0$ is dimensionless, and $\tau\rightarrow0$ is an infinitesimal amount of time (one could also write $\dot\Gamma$ as $\dot\Gamma(t) = \gamma_0\delta(t-t_h)$ with $\Gamma=\gamma_0/\tau\rightarrow\infty$). Upon integration across $t=t_h$, we find, from equ.(\ref{alpha}) and (\ref{jumpcondition}), that $\zeta_m^+=\zeta_m^+-\zeta_m^-\approx-\gamma_0(\zeta_m^- +\Phi)/(12H\tau)$.

Since both $\zeta_m^- + \Phi$ and $\tau$ are infinitesimally small, $\zeta_m^+$ is an indeterminate of the the form $0/0$, and can certainly assume the observed value of $3\times10^{-5}$. The subsequent evolution of $\zeta_v$ is, by equ.(\ref{npair}), on the time scale of $1/H$ ($\Gamma=\gamma_0/\tau\rightarrow\infty$ is $\gg H$). Yet, by the first of the three terms on the right side of equ.(\ref{coherent}) with $\rho_m^+\approx0$, $\rho_m^+\approx\rho_m^-$ (since $\Gamma\rightarrow\infty$, the decay of particles in radiation is effectively instantaneous; also note that the $\dot\epsilon$ term vanishes because $\epsilon=1+w_m=1$ for massive particles), $\zeta$ rapidly converges to become $\zeta_v$ on the timescale $\approx 4/\Gamma$.
Moreover, the final and constant $\zeta$ applies to all superhorizon modes, as $\gamma_0(\zeta_m^-+\Phi)/\tau$ is the same finite quantity resulting from the spontaneous decay process, which is a universal constant characteristic only of the process, and is independent of the wavelength of a perturbation mode.

\section{Conclusion: Scale-Invariant Adiabatic Curvature Modes without Inflation}
We demonstrated by solving the coupled linear equation that 'isocurvature' modes, generated during the reheating decay of the massive scalar field into radiation, can alter the adiabatic curvature  mode amplitude if a large decay rate $\Gamma$ into photons arises suddenly as a step function in time, such that $\dot\Gamma=\Gamma\delta(t-t_h)$. 

Although $\zeta$ does settle to another constant (i.e. $\zeta$ is an invariant) in the end, for reheating models of inflation any quantum fluctuations successfully propagating into the reheated thermal Universe remains precarious because of the large spikes in $\zeta(t)$ at times during coherent oscillations when $\epsilon = 1+w_v$ vanishes. 
Here we adopted ratios of $\omega_h/\Gamma$ between $100$ and $10^4$, but the massive scalar field at the end of inflation (and before reheating) could cause the frequency $\omega_h$ of coherent oscillations to reach values $\gg10^4\omega_h$. Under the scenario, $\zeta(t)$ could become $>1$, and perturbation theory would break down.

\section*{References}
\bibliographystyle{jphysicsB}
\bibliography{ReheatingCQG}

\appendix
\section{Mathematics Supplement}
\label{appendix:MathSupplement}
This supplement contains the details on the derivation in the main body of the paper.
Let's start with the linear isotropic perturbations on a spatially-flat FRW universe, which is commonly written as 
\begin{equation}
ds^2 = -(1+2\Psi)dt^2+ 2aB_{\,,i}dtdx^i + a^2[(1-2\Phi)\delta_{ij}+2E_{\,,ij}]dx^idx^j,
\end{equation}
where $a(t)$ is the scale factor. $\Psi$, $\Phi$, $B$ and $E$ are scalar perturbations. $E$ and $B$ are not involved in the discussion in the current paper.  
The total perturbation on the background, referring to equ.(\ref{total}), is 
    $\zeta = -\Phi -H\delta\rho/\dot\rho$. 
In the long-wavelength limit, the divergence of the momenta of the zero-shear gauge is negligible. The change of the perturbation on the total density, from the energy conservation, is 
\begin{equation}
    \dot{\delta\rho}  = -3H(\delta\rho + \delta p) + 3(\rho + p)\dot\Phi.\label{dotrho}
\end{equation}
Taking the time derivative of $\zeta$
\begin{equation*}
    \dot\zeta =-\dot\Phi -\dot H\frac{\delta\rho}{\dot\rho} - H\frac{\dot{\delta\rho}}{\dot\rho} + H\frac{\delta\rho \ddot\rho}{\dot\rho^2}
\end{equation*}
and substituting equ.(\ref{trates}), (\ref{dotrho}) and $H^2=\rho$ into the expression of $\dot\zeta$, one gets the equation of $\dot\zeta$ in the multi-component fluid,
\begin{equation}
    \dot\zeta = - \frac{H}{\rho + p}~(\delta p-c_s^2\delta\rho)
    = \fr{-1}{6\dot\rh(\rh+p)} \sum_{\mu,\nu} \dot\rh_\mu \dot\rh_\nu (w_\mu - w_\nu) S_{\mu \nu},\label{DotZeta1}
\end{equation}
where $S_{\mu\nu}=3(\zeta_\mu-\zeta_\nu)$ describes the relative entropy between two different components. 
On the right-handed side of the second equal sign, we assign $\mu=v$ for the curvature and $\nu=r$ for the radiation. 
One obtains the change of $\zeta$ in the two components fluids, which is
\beq
\dot\zeta = -\frac{H}{\dot\rho^2}\dot\rho_v\dot\rho_r(w_v-w_r)S_{vr}.
\label{DotZeta2}
\eeq

The unperturbed equation of energy conservation in the $\mu$ component is
\begin{equation}
\dot\rho_\mu = -3H(\rho_\mu + p_\mu) + Q_\mu,
\label{muComponent}
\end{equation}
where $Q_\mu$ is the energy transfer between components.
When the curvature converts to the radiation in a rate of $\Gamma$ during the reheating stage, we obtain equ (\ref{rates_v}) for $\mu=v$ by setting $Q_\mu=-\Gamma\rho_v$ and obtain equ (\ref{rates_r}) for $\mu=r$ when $Q_\mu=\Gamma\rho_v$.

The first order of the perturbation of equ.(\ref{muComponent}) in the the $\mu$-component, in the long-wavelength limit \cite{MWU2003,Riotto2003}, is 
\begin{equation}
    \dot{\delta\rho}_\mu +3H(\delta\rho_\mu + \delta p_\mu) 
    = 3(\rho_\mu + p_\mu)\dot\Phi + Q_\mu\Psi + \delta Q_\mu.
    \label{DeltamuComponent1}
\end{equation}
where $\delta Q_\mu$ is the perturbation of  the energy transfer. 
The $00$ component of the perturbed Einstein equation for the large wavelength (or super-horizon) version is 
\beq
\dot\Phi +H\Psi = -\frac{H\delta\rho}{2\rho}.\label{dotPhi}
\eeq
Combining equ.(\ref{muComponent}) and (\ref{dotPhi}), we re-arrange equ.(\ref{DeltamuComponent1}) in terms of $\zeta_\mu$, defined by equ.(\ref{zenu}), to solve the change of $\zeta$ of the $\mu$-component,
\beq
\dot\zeta_\mu
= \frac{3H^2}{\dot\rho_\mu}(\delta p_\mu - c_\mu^2 \delta\rho_\mu)
-\frac{H}{\dot\rho_\mu}\Bigg(\delta Q_\mu - \frac{\dot Q_\mu \delta\rho_\mu}{\dot\rho_\mu}
+\frac{Q_\mu\dot\rho\delta\rho_\mu}{2\rho\dot\rho_\mu} - \frac{Q_\mu\delta\rho}{2\rho}\Bigg).
\label{DotZetaMu}
\eeq
When energy transfers from the curvature to the radiation in a constant rate of $\Gamma$, there are $\delta Q_v = -\Gamma\delta\rho_v$, 
$\delta Q_r=\Gamma\delta\rho_v$,
$\dot{Q_v}=-\Gamma\dot\rho_v$, 
and $\dot{Q_r}=\Gamma\dot\rho_r$.
Substituting these relations into equ.(\ref{DotZetaMu}), one immediately calculates that 
\beq
\dot\zeta_v = -\frac{\Gamma}{6}\frac{\rho_v}{\rho}\frac{\dot\rho_r}{\dot\rho_v}S_{vr},
\ {\rm and}~
\dot\zeta_r = \frac{\Gamma}{3}\frac{\dot\rho_v}{\dot\rho_r}\Big(1-\frac{\rho_v}{2\rho}\Big)S_{vr}.
\label{DotZetaVR}
\eeq
Next, the change of $S_{vr}$ in the fluid with two components can be yielded from $\dot{S_{vr}}=3(\dot\zeta_v-\dot\zeta_r)$
\beq 
\dot S_{vr} = \Ga\left[\fr{\rh_v}{2\rh} \fr{\dot\rh_v}{\dot\rh_r} \left(1-\fr{\dot\rh_r^2}{\dot\rh_v^2}\right) - \fr{\dot\rh_v}{\dot\rh_r}\right]S_{vr}. \label{Seq2} 
\eeq

Moreover, because of the absence of the anisotropic stress, namely the energy-momentum tensor is diagonal, one concludes, from the non-diagonal component of Einstein equation, that $\Phi=\Psi$ \cite{Riotto2003,Brandenberger2004}. Substitute this condition into equ.(\ref{dotPhi}), equ.(\ref{Poisson}) and (\ref{nPoisson}) are yielded, namely
\begin{equation}
    \dot\Phi + H\Phi = \frac{\dot\rho}{2\rho}(\zeta+\Phi)
    = \frac{\dot\rho_v(\zeta_v+\Phi)+\dot\rho_r(\zeta_r + \Phi)}{2\rho}.
\label{dotPhi2}
\end{equation}

\section{Solution to Superhorizon Growth Equations at and after the Commence of Reheating}
\label{appendix:SolutionReheating}
Among the three coupled growth equations (\ref{dotze4}), (\ref{npair}), and (\ref{nPoisson}) for $\zeta$, $\zeta_v$, and $\Phi$, respectively, only $\zeta_v$ jumps across the boundary of $t=t_h$ because, in equ.(\ref{npair}), $\dot\zeta_v$ has a term of $\dot{\Gamma}=\Gamma\delta(t-t_h)$. 
Integrating the term across $t=t_h$ and applying equ.(\ref{rates_v}) yields, in the limit of $\Gamma\gg H$,
\begin{equation}
\zeta_v(t_h^+) = -\alpha \zeta_v(t_h^+) + (1-\beta)\zeta_v(t_h^-) - \beta\Phi(t_h),
\label{zvthplus}
\end{equation}
where the factor $\alpha$ is assigned to as
\begin{equation}
\alpha = \frac{\Gamma}{6[1+w_v(t_h)]H(t_h)+\Gamma};
\ 
\eta = \frac{1}{6[1+w_v(t_h)]H(t_h)};
\ 
\beta = \eta\Gamma.
\label{alpha}
\end{equation}
Since neither $\zeta(t)$ nor $\Phi(t)$ jump across $t=t_h$, this leads to the jump condition for $\zeta-\zeta_v$ from $\zeta^--\zeta_v^-=0$ ($\zeta^-=\zeta(t_h^-)$ etc.) to 
\begin{equation}
\zeta^+-\zeta_v^+ = \zeta_v^--\zeta_v^+ = \frac{\alpha+\beta}{\alpha + 1} \zeta_v(t_h^-) + \frac{\beta}{\alpha+1}\Phi(t_h).
\label{jumpcondition}
\end{equation}
It is obvious that $\alpha\simeq1$ and $\beta\gg1$ when not a necessary condition.
The jump condition is approximated to $ \zeta_v^--\zeta_v^+\approx [\zeta_v(t_h^-)+\Phi(t_h)]\beta/2 = \Gamma[\zeta_v(t_h^-)+\Phi(t_h)]\eta/2$.
Because $\Gamma=0$ when $t<t_h$, $\Phi(t_h)$ is independent of $\Gamma$ at $t=t_h$, so is $\zeta_v(t_h^-)$. Then $\zeta_v^--\zeta_v^+$ is approximately proportional to $\Gamma$ in the slope as $[\zeta_v(t_h^-)+\Phi(t_h)]\eta/2$.
Substituting equ.(\ref{jumpcondition}) into (\ref{dotze4}) now leads to a finite $\dot\zeta$ after reheating commence, given by
\begin{equation}
\dot{\zeta}^+ = 
\frac{3H(t_h)\beta}{2\dot\rho(t_h)}
\bigg\{\Big[w_v(t_h)-w_r\Big]\dot\rho_v(t_h^+)+\rho_v(t_h)\dot{w_v}(t_h)\bigg\}
\Big[\zeta(t_h^-) + \Phi(t_h)].
\label{zthplus} 
\end{equation}
This is the origin of the slope in $\zeta(t)$ immediately after $t=t_h$.
 
Equ.(\ref{nPoisson}), by following which $\Phi(t)$ evolves from its initial value $\Phi_0$ to $\Phi(t_h)$ at the instance of $t=t_h$, makes a one-to-one mapping between $\Phi_0$ and $\Phi(t_h)$. So when $\Phi_0$ is in a Gauss distribution $\mathcal{N}(0,\sigma_{\Phi0})$, the distribution of $\Phi(t_h)$ is Gaussian as well, referred to as $\mathcal{N}(0,\sigma_{\Phi h})$. Meanwhile, when $t<t_h$, $\zeta_v$ keeps in a constant, which makes $\zeta_v^-$ equal to its initial value. If the initial value $\zeta_{v0}$ is in a Gauss distribution $\mathcal{N}(0,\sigma_{\zeta0})$, $\zeta_v^-$ will share the same distribution along with $\zeta_{v0}$. 
We also notice that because $\zeta_v$ keeps itself in the constant and $\zeta(t)$ approaches $\zeta_v$ at $t\gg1$, there is $\zeta_T = \zeta_v^+$. $\zeta_v^+$ is solved from equ.(\ref{zvthplus}) trivially as 
\begin{equation}
    \zeta_T = \zeta_v^+ = \frac{1-\beta}{1+\alpha}\zeta_v^- -\frac{\beta}{1+\alpha}\Phi(t_h) 
    \approx -\frac{\beta}{2}[\zeta_v^- +\Phi(t_h)]. 
    \label{zvthplus2}
\end{equation}
Here $\zeta_v^-$ and $\Phi(t_h)$ are both Gaussian, namely $\zeta_v^-\sim\mathcal{N}(0,\sigma_{\zeta0})$ and $\Phi(t_h)\sim\mathcal{N}(0,\sigma_{\Phi h})$, respectively,  and independent of each other.  $\zeta_T$, as a linear combination of two Gaussian variables, must be Gaussian.  Its standard deviation $\sigma_T$ is evaluated as $\sigma_T\approx \sqrt{\sigma_{\zeta0}^2 + \sigma_{\Phi h}^2}\cdot\beta/2$. Because factor $\beta$ is proportional to $\Gamma$, we obtain $\sigma_T\propto\Gamma$.

\section{Post-Reheating \texorpdfstring{$(\dot\Gamma=0)$} Solution to the Growth Equations for Early and Late Times}
\label{appendix:PostReheating}
At time $t\geq t_h$, one can subtract equ.(\ref{npair}) from (\ref{dotze4}) to get 
\beq 
\fl
\dot\ze - \dot\ze_v = -\left[(\frac{4}{3}-\ep)\fr{(3\epsilon H+\Ga)\rh_v}{\ep\rh_v + \frac{4}{3}\rh_r} +\fr{\rh_v}{\ep \rh_v + \frac{4}{3} \rh_r} \dot\ep  + \fr{\Gamma(3\ep\rh_v + 4\rh_r)}{2\rh(3\epsilon + \Gamma/H)}\right](\ze - \ze_v);~{\rm for}~t>t_h.
\label{coherent_app} 
\eeq 
In the limit of $\omega\gg\Gamma\gg H$ and for these early times the middle term of (\ref{coherent}) dominates, so that 
\begin{equation}
\fl
\zeta(t) - \zeta_v(t) = \frac{1-\cos(\sqrt{2\epsilon_h})}{1-\cos[\omega_h(t-t_h)+\sqrt{2\epsilon_h}~] + \frac{4\rho_r}{3\rho_v}}
\bigg[\zeta(t_h^+)-\zeta_v(t_h^+)\bigg], ~ t\geq t_h^+.
\label{Dzeta}
\end{equation}
Thus, in these early times, $\zeta-\zeta_v$ oscillates and exhibits sharp spikes at times when the denominator is minimized at $4\rho_r/(3\rho_v) \simeq 4\Gamma(t-t_h)/3$.
Evidently, then the larger the ratio of $\omega_h$ to $\Gamma$ (or in the case of a fixed $\omega$ but $\Gamma$ decreasing), the taller the spikes. This is borne out by the numerical solution (namely, Section \ref{sec::numerical}) of the context in the main body.

On the other hand, at late times $t\gg1/\Gamma$ when $\rho_v\rightarrow0$ and $\rho\rightarrow\rho_r$, the last term on the right side of (\ref{coherent}) dominates, and 
\begin{equation}
\dot\zeta - \dot\zeta_v = -2H(\zeta-\zeta_v) = - \frac{1}{t}(\zeta-\zeta_v),
\label{Dzeta2}
\end{equation}
or $\zeta-\zeta_v$ tends to zero as $1/t$. By (\ref{npair}), $\zeta_v$ tends to a constant, and hence $\zeta$ reaches the same constant at large $t$.

\end{document}